\documentstyle[12pt,aps]{revtex}

\title{Nuclear magnetic resonance spectrum of  ${}^{31}P$ donors in silicon quantum computer}
\author{A.A. Larionov, L.E. Fedichkin,  A.A. Kokin, K.A. Valiev}
\address{Institute of Physics and Technology, Russian Academy of Sciences, 
34, Nakhimosky pr., Moscow, 117218, Russia\\
e-mail: qubit@ftian.oivta.ru}
\begin{document}
\draft\onecolumn
\maketitle

\begin{abstract}
The influence of the electric field created
by a gate potential of the  silicon quantum computer
on the hyperfine interaction constant (HIC) is obtained.
The errors due to technological inaccuracy of location of donor atoms
under a gate are evaluated.
The energy spectra of electron-nuclear spin system of two interacting
donor atoms with various values of HIC are calculated.
The presence of two pairs of anticrossing levels in the ground
electronic state is shown. Parameters of the structure at which errors
rate can be greatly minimized are found.
\end{abstract}

\section{Introduction}
The first approach to the NMR quantum computer realization simultaneously
proposed in 1997 by two groups of investigators \cite{Chuang,2}, and then
experimentally confirmed \cite{3,4}, consists in using organic liquids,
where different molecules with a few connected with each other nonequivalent
nuclear spins-qubits function as independent quantum computers.
Another approach, which is qualitatively different from the bulk-ensemble one
and is still unrealized, was proposed and analyzed in details by Kane \cite{5}
and elaborated in \cite{8,13,14}. It is based on creating artificial multi-spin
system and using individual addressing to different spins-qubits. For
this purpose it is suggested to use silicon-based structure of MOS-type,
where donor atoms of stable phosphorus isotope ${}^{31}P$ are implanted into
the thin layer of spinfree silicon isotope ${}^{28}Si$ at a definite depth.
Donor atoms replace silicon ones in knots of crystal lattice. Such a donor
has shallow impurity states, which have a large magnitude of the effective
Bohr radius and nuclear spin $I=1/2$. Every donor atom with nuclear spin in
semiconductor structure is supposed to be placed regularly with adequate
accuracy under its "own" control metallic gate (gate A), which is separated
from the silicon surface by a thin dielectric layer (for example, silicon
oxide with the thickness of about some nanometers). A-gates form linear
regular structure of arbitrarily length with period l (Fig. 1).
Changing of the electric potential of J-gates, which are placed
between A-gates, allows us by redistributing electron density between
neighboring donors to control the degree of overlapping of  electron wave
functions, which are localized on neighboring donors a and b, and to control
the constant of exchanging interaction J, and the constant of scalar
interaction of their nuclear spins ${I}_{ab}$ as well.
It is assumed that with the help of the electric field, induced by A-gates,
one can change the distribution of electronic density near the nuclei in
the ground state by choosing individually, correspondingly, each donor
atom nuclear spin resonance frequency, which is determined by the hyperfine
interaction with its electron spin. This allows to fulfill quantum operations
by the way of selective reaction of resonance radio-frequency pulses on
nuclear spins of the given donors. The second section of the work studies
the influence of the electric field, created by potential V on the A-gate,
on hyperfine interaction constant (HIC) of the donor atom in two cases:
for gates having the form of a round disc and that of a strip. In the
third section HIC fault approximation is made, which is determined by
the technological placement inaccuracy of donor atom under the gate.
In the fourth section the results of full calculation of the
electron-nuclear spin system energy spectrum of two interacting donor
atoms with different magnitude of HIC are demonstrated. Separately the
phenomenon of hyperfine energy levels crossing and anti-crossing is
investigated, owing to which the information about nuclear spin state
is transferred to the electron spin of donor atom when moving through the
point of crossing. This property was suggested to be used in the process
of electrical measurement of a definite nuclear spin state \cite{5,11}.
It is also very important when using magnetically-mechanic resonance method
\cite{12}.

\section{Dependence of the hyperfine interaction constant
on the electrical field}

Spin Hamiltonian of the hyperfine interaction is:

\begin{equation} \label{R1}
\hat H_{IS}  = A \left( \hat{\vec{I}} \hat{\vec{S}} \right),\,A\left( V \right) = \frac{{8\pi }}{3}\left| {\Psi _0 \left( {0,V} \right)} \right|^2 2\mu _B g_N \mu _N  \cdot \left( {\frac{{\mu _0 }}{{4\pi }}} \right)
\end{equation}

where $ \mu _N  = 5.05 \cdot 10^{ - 27} $ J/T --- nuclear magneton,
$\mu _0  = 4\pi  \cdot 10^{ - 1} $ $T^{2}sm^{3}/J$, $g_N  = 2,26$  --- Lande's
factor for ${}^{31}P$, $\mu _B  = 9.27 \cdot 10^{ - 24} $ J/T --- Bohr magneton,
 and $\Psi _0 \left( {0,V} \right)$  --- wave function of the electron in the
 ground state placed on the nuclei. Experiments give the value of the
 probability density of the electron to be on the nuclei
 $\left( {\Psi _0 \left( 0 \right)} \right)^2  = 0.43 \cdot 10^{24} $
$sm^{-3} $ at V=0 \cite{6}.

Taking into account that intrinsic semiconductor at low temperatures
 behaves like dielectric and supposing that $ D >  > l_A \sim c >  > d$
 (Fig. 1), we express the electric field on the donor atom depending
 on potential V on gate A, (round disc with radius $ a = \frac{{l_A }}{2}$),
which lies on the surface of the dielectric. To find the electric
potential induced by the round disc at the point with the coordinates
$\vec r' = \left( {\rho',z'} \right),\;\rho '^2  = x'^2  + y'^2$, we
use known expression \cite{7}:

\begin{equation} \label{R2}
\varphi \left( {\rho ',z'} \right) = \frac{{2V}}{\pi }arctg\sqrt {\frac{{2a^2 }}{{\rho '^2  + z'^2  - a^2  + \sqrt {\left( {\rho '^2  + z'^2  - a^2 } \right)^2  + 4a^2 z'^2 } }}}
\end{equation}

On the line, which passes through the gate center $
\rho ' = \sqrt {x'^2  + y'^2 }  = 0
$
 , near the donor atom ($z'=c$), we receive:

\begin{equation} \label{R3}
\varphi \;\left( {\rho ',z'} \right) = \frac{{2V}}{\pi }arctg\frac{a}{c} - E_c \left( {z' - c} \right) + \frac{{E'_c }}{2}\left( {z' - c} \right)^2  - \frac{{E''_c }}{2}\rho '^2  +  \ldots ,
\end{equation}

where

\begin{equation} \label{R4}
E_c  = \frac{{2V}}{\pi } \cdot \frac{a}{{a^2  + c^2 }},{\kern 1pt} E'_c  = \frac{{4V}}{\pi } \cdot \frac{{ac}}{{\left( {a^2  + c^2 } \right)^2 }},{\kern 1pt} E''_c  = \frac{{4V}}{\pi }\sqrt 2 \frac{{ac^4 }}{{\left( {a^2  + c^2 } \right)^{7/2} }}
\end{equation}

the electric field, its gradient in the $z'$ axis direction and the electric
field gradient along the radial direction near the donor atom.
When crystalline lattice and external electric field are absent,
the wave function of this state is trans-formed into the hydrogen-like
function, which corresponds to 1S state. Shroedinger equation for this
function in the electric field must be complemented by the perturbation
operator $
\Delta \hat H = e \cdot \varphi \,\left( {\rho ',z'} \right)
$
 ( $
e = 1.6 \cdot 10^{ - 19}
$
C --- elementary charge) \cite{8}. Using (\ref{R3}), (\ref{R4}) in x, y, z basis
(Fig. 1) we receive the following expression ($
\rho ^2  = x^2  + y^2
$
 ):

\begin{equation} \label{R5}
\Delta \hat H =  - \frac{{2Ve}}{\pi }arctg\frac{a}{c} + eE_c  \cdot z - \frac{{eE'_c }}{2} \cdot z^2  + \frac{{eE''_c }}{2} \cdot \rho ^2  = \Delta \hat H_0  + \Delta \hat H_1  + \Delta \hat H_2  + \Delta \hat H_3
\end{equation}

The wave function $
F_j \left( {\vec r} \right)
$
 correction, specified by the perturbance (\ref{R5}), is:

\begin{equation}  \label{R6}
F_j  = F_j^{\left( {1s} \right)}  + \Delta F_{j,1}^{\left( {1s} \right)}  + \Delta F_{j,2}^{\left( {1s} \right)} ,
\end{equation}

where

\begin{equation}   \label{R7a}
\Delta F_{j,1}^{\left( {1s} \right)}  = \frac{{\Delta H_{2s,1s} }}{{\delta \;E}}F_j^{(2s)}
\end{equation}

- the first order correction of the perturbation theory, $
\Delta H_{2s,1s}  = \int {d^3 r}  \cdot \left( {F_j^{\left( {2s} \right)}
\left( r \right)} \right)^ *  \left(
{\Delta \mathord{\buildrel{\lower3pt\hbox{$\scriptscriptstyle\frown$}}
\over H} } \right)F_j^{\left( {1s} \right)} \left( r \right)
$
  --- matrix element of the perturbation operator and $
\delta E = E_{1s}  - E_{2s}
$
  --- energy residual of unperturbed donor states, and

\begin{equation} \label{R7b}
\Delta F_{j,2}^{\left( {1s} \right)}  =  - \frac{{F_j^{\left( {1s} \right)} }}{2}\sum\limits_{m \ne (1s)} {\frac{{\left| {\Delta H_{m,1s} } \right|^2 }}{{\left( {E_{1s}  - E_m } \right)^2 }}} ,\;m = 2p,3p, \ldots
\end{equation}

- the second order correction (here $
E_m
$
  --- unperturbed energy levels of donor,  $
\Delta H{}_{m,n}
$
 --- matrix elements of the perturbation operator). During further
calculations we shall use hydrogen-like functions as unperturbed wave
functions. They are independent of j \cite{10}:

\begin{equation}   \label{R8}
F^{(1s)} \left( r \right) = \frac{1}{{\sqrt \pi   \cdot \left( {a^ *  } \right)^{3/2} }}e^{ - \frac{r}{{a^ *  }}} ;\,F^{(2s)} \left( r \right) = \frac{1}{{4\sqrt {2\pi }  \cdot \left( {a^ *  } \right)^{3/2} }}\left( {2 - \frac{r}{{a^ *  }}} \right)e^{ - \frac{r}{{2a^ *  }}}
\end{equation}

where $
r^2  = x^2  + y^2  + z^2
$
 (Fig. 1), and  $
a^*  = \frac{{4\pi \varepsilon \varepsilon _0 \hbar ^2 }}{{m^* e^2 }} = 20\mathop A\limits^0
$
 --- effective Bohr radius \cite{9}, $
\varepsilon  = 11.9
$, $
\varepsilon _0  = 8.85 \cdot 10^{ - 14}
$
 F/sm --- vacuum permittivity, $
\hbar  = 6.62 \cdot 10^{ - 34}
$
 J*sec, $
m^*  = 0.31m_e
$, $
m_e  = 9.1 \cdot 10^{ - 31}
$
kg. The second order terms in (\ref{R6}) give an expression a the
respective correction of the hyperfine interaction constant of the
donor atom:

\begin{equation} \label{R9}
\frac{{\Delta A\left( V \right)}}{A} \equiv \frac{{\left( F \right)^2  - \left( {F^{\left( {1s} \right)} } \right)^2 }}{{\left( {F^{\left( {1s} \right)} } \right)^2 }} \cong 2\frac{{\Delta F_1^{\left( {1s} \right)} }}{{F^{\left( {1s} \right)} }} + 2\frac{{\Delta F_2^{\left( {1s} \right)} }}{{F^{\left( {1s} \right)} }} + \left( {\frac{{\Delta F_1^{\left( {1s} \right)} }}{{F^{\left( {1s} \right)} }}} \right)^2
\end{equation}

The second term in (\ref{R9}), according to \cite{8,14}, may be evaluated as:

\begin{equation} \label{R10}
\frac{{2\Delta F_2^{(1s)} }}{{F^{(1s)} }} =  - \frac{9}{4} \cdot 4\pi \varepsilon _0 \left( {a^* } \right)^3  \cdot \frac{{E_c^2 }}{{\Delta E}}
\end{equation}

where $
\left( {\Delta E} \right)^{ - 1}
$
  --- some mean value of $
\left( {E_m  - E_{1s} } \right)^{ - 1}
$
 . At $
\Delta E \sim 0.04
$
 ev, $c=2a=10$ nm it was obtained that (V in volts) $
\frac{{2\Delta F_2^{(1s)} }}{{F^{(1s)} }} =  - 0.19V^2
$
 .

When calculating $
\Delta H_{2s,1s}
$
, we shall take into account that the item in the perturbance $
\Delta \hat H_0
$
 does not contribute because of orthogonality of functions $
F^{\left( {1s} \right)} \left( {\vec r} \right)
$
 and  $
F^{\left( {2s} \right)} \left( {\vec r} \right)
$
, and $
\Delta \hat H_1
$
 does not contribute because of oddness of the function by $z$ under integral.
 Resulting,

\begin{equation} \label{R11}
\Delta H_{2s,1s}  = \frac{{2^8 \sqrt 2 }}{{3^6 }}eE'_c \left( {a^* } \right)^2  - \frac{{7^2 2^5 }}{{3^6 \sqrt 2 }}eE''_c \left( {a^* } \right)^2
\end{equation}

Taking into account that  $
\frac{{F^{(2s)} }}{{F^{(1s)} }} = \frac{{\sqrt 2 }}{4}
$
, (\ref{R9}) appears as:

\begin{equation} \label{R12}
\frac{{\Delta A\left( V \right)}}{A} \cong  - 9\pi \varepsilon _0 \left( {a^* } \right)^3 \frac{{E_c^2 }}{{\Delta E}} + \frac{{2^8 eE'_c \left( {a^* } \right)^2 }}{{3^6 \delta \,E}}\left( {1 - \frac{{7^2 E''_c }}{{2^4 E'_c }}} \right) + \left( {\frac{{2^7 }}{{3^6 }}\frac{{eE'_c \left( {a^* } \right)^2 }}{{\delta \,E}}\left( {1 - \frac{{7^2 E''_c }}{{2^4 E'_c }}} \right)} \right)^2
\end{equation}

The energy residual $
\delta E
$
 can be evaluated, if we use the formula for hydrogen-like atoms  $
\delta E =  - \frac{3}{8} \cdot \frac{{e^2 }}{{4\pi \varepsilon \varepsilon _0 a^* }}
$
. The evaluation for silicon gives $
\delta E =  - 0.023
$
 ev. Using the above-mentioned evaluations with the help of (\ref{R12})
for $c=2a=10$ nm  we shall result:

\begin{equation} \label{R13}
\frac{{\Delta A\left( V \right)}}{A} \cong 0.55V - 0.09V^2
\end{equation}

Let us consider the case, when the A-gates have a form of infinitely
long strips with the width  $
2a = l_A
$
. They are placed at the distance D from the conducting substrate,
from which the gate potential is counted out. Conducting similiar
operations we receive for the donor atom HIC the following expression
($c=2a=10$ nm, $
\frac{D}{a} = 100
$
):

\begin{equation}  \label{R14}
\frac{{\Delta A\left( V \right)}}{A} \cong  - 0.063V^2
\end{equation}

Voltage in (\ref{R14}) is expressed in volts, as written above.

\section{Influence of the technological spread of donor atoms on the HIC.
Voltage error on a gate}

Technological inaccuracies of the donor atom place under the gate
relatively $x=0$, $z=c$ are designated $\delta x$
 and $\delta z$
. We investigate only the strip gate variant:
Factorizing potential expression for strip gate by degrees of $\delta x$
 and $\delta z$
, we write it as:

\begin{equation} \label{R15}
\varphi \left( {x,z} \right) \approx \varphi \left( {0,c} \right) + \frac{{\partial \varphi \left( {\delta x,c + \delta z} \right)}}{{\partial z}}\left( {z - c} \right) +  + \frac{1}{{2!}}\frac{{\partial ^2 \varphi \left( {\delta x,c + \delta z} \right)}}{{\partial z^2 }}\left( {z - c} \right)^2  + \frac{{\partial \varphi \left( {\delta x,c + \delta z} \right)}}{{\partial \left( {x^2 } \right)}}x^2
\end{equation}

where

\begin{equation} \label{R16}
\begin{array}{l}
 \frac{{\partial \varphi \left( {\delta x,c + \delta z} \right)}}{{\partial z}} =  - \bar E_c \left\{ {1 - \frac{{\delta z}}{{c\left( {1 + {{a^2 }}/{{c^2 }}} \right)}} - \frac{{\left( {\delta x} \right)^2 \left( {2c^2  - a^2 } \right)}}{{2\left( {a^2  + c^2 } \right)^2 }}} \right\}; \\
 \frac{{\partial ^2 \varphi \left( {\delta x,c + \delta z} \right)}}{{\partial z^2 }} = \bar E'_c \left\{ {1 - \frac{{\delta z\left( {2c^2  - a^2 } \right)}}{{c\left( {a^2  + c^2 } \right)}} - \frac{{\left( {\delta x} \right)^2 \left( {4c^4  + a^2 c^2  - a^4 } \right)}}{{2c^2 \left( {a^2  + c^2 } \right)^2 }}} \right\}; \\
 \frac{{\partial \varphi \left( {\delta x,c + \delta z} \right)}}{{\partial \left( {x^2 } \right)}} =  - \frac{1}{2}\bar E''_c \left\{ {1 - \frac{{\delta z\left( {2c^2  - a^2 } \right)}}{{c\left( {a^2  + c^2 } \right)}} - \frac{{\left( {\delta x} \right)^2 \left( {2c^2  + a^2 } \right)}}{{2\left( {a^2  + c^2 } \right)^2 }}} \right\}. \\
 \end{array}
\end{equation}

We consider $
\delta z \sim \left( {\delta x} \right)^2
$
, that is why corrections of  $
\left( {\delta z} \right)^2
$
 degree are omitted being of the next order. Using expressions (\ref{R12}),
 (\ref{R14})--(\ref{R16}), we receive the final expression for the relative
error of the HIC:

\begin{equation} \label{R17}
\frac{{\delta A}}{A} = \delta z\left\{ {0.063V^2 \frac{{2c}}{{a^2  + c^2 }}} \right\} + \left( {\delta x} \right)^2 \left\{ {0.063V^2 \frac{{2c^2  - a^2 }}{{\left( {a^2  + c^2 } \right)^2 }} - 0.085V\frac{{2c^4  - a^4 }}{{2c^2 \left( {a^2  + c^2 } \right)^2 }}} \right\}
\end{equation}

The second item in (\ref{R17}) nullifies in a variety of values of $a$, $c$
and $V$. That is why by defining necessary relative error it is possible
to determine structure parameters a and c and voltage V, when the needed
accuracy is realized. One percent relative error of the hyperfine
interaction constant corresponds to the donor atom deviation along the $z$
axis, which is connected with the technological inaccuracy of
placing in 2-3 nm.

Voltage error on the gate, which is admissible during correctly
conducted quantum calculation process, can be determined from the
final expressions for HIC (\ref{R13}) and (\ref{R14}). The strip width
of radio frequency pulses must be at
least one order smaller than the resonant frequency of nuclear spins,
which is about hundreds of kHz. Considering the HIC value, which corresponds
to the strip width of resonant pulse
$
\delta \left( {\Delta A} \right)\sim 10^4
$
 Hz, for the gate voltage error we receive $10^{ - 4}  \div 10^{ - 3} $V.
It is worth of noting that in the case of round discs the distinguished
voltage exists, that leads to the significant increasing of the admissible
voltage error. It is caused by the presence of the linear by V term
in (\ref{R13}).

\section {Electron-nucleus spin system energy spectrum of two
interacting donor atoms}

Consider two donor atoms separated from each other by distance l. This distance must satisfy the following condition: constant J of the electrons' effective exchange interaction, determined by the partial overlap of their wave functions, must correspond to the possibility of making two-qubit operations. Spin Hamiltonian of this system is:

\begin{equation} \label{R18}
\begin{array}{l}
 \hat{H} = 2\mu _B \vec{B}\left( \hat{\vec{S_a}}  + \hat{\vec{S_b}} \right) +
 J\left( \hat{\vec{S_a}} \hat{\vec{S_b}} \right) + \Delta \hat{H}
 = \hat{H_0}  + \Delta \hat{H}, \\
 \Delta \hat{H} =  - g_N \mu _N \vec{B}\left( \hat{\vec{I_a}}  + \hat{\vec{I_b}} \right) + A_a \left( \hat {\vec {I_a}}
\hat{\vec{S_a}} \right) + A_b \left( \hat {\vec {I_b}} \hat{\vec {S_b}}
\right), \\
 \end{array}
\end{equation}

where  $
\hat{\vec{S_a}}
$
,  $
\hat {\vec{I_a}}
$
,  $
\hat {\vec{S_b}}
$
,  $
\hat {\vec{I_b}}
$
 - spin operators of the electron and the nuclei for the first and
the second atom,  $
\vec{B}
$
 --- the magnetic field (it is directed in parallel to $z$ axis
(see Fig. 1)),  $
A_a
$
 and  $
A_b
$
 --- hyperfine interaction constants depending, generally speaking,
on gate potentials,  $
\mu _B
$
 --- Bohr magneton,  $
\mu _N
$
 --- nuclear magneton,  $
g_N  = 2.26
$
 --- Lande factor for  $
{}^{31}P
$
.

We consider $
\left| {M_a ;M_b ;m_a ;m_b } \right\rangle
$
 to be the basis (eigen states in the strongest fields), where $
M_a
$
,  $
M_b
$
,  $
m_a
$
,  $
m_b
$
 --- quantum numbers of the electrons' and nuclei's spin projections.
Designating them by arrows up $\left(  \uparrow  \right)$
 and down $\left(  \downarrow  \right)$
, we have (numbers are strings or rows in the matrix):

\begin{equation}
\begin{array}{l}
 \left| 1 \right\rangle  = \left| { \uparrow  \uparrow  \uparrow  \uparrow } \right\rangle \quad \left| 5 \right\rangle  = \left| { \uparrow  \downarrow  \uparrow  \uparrow } \right\rangle \quad \left| 9 \right\rangle  = \left| { \downarrow  \uparrow  \uparrow  \uparrow } \right\rangle \quad \left| {13} \right\rangle  = \left| { \downarrow  \downarrow  \uparrow  \uparrow } \right\rangle  \\
 \left| 2 \right\rangle  = \left| { \uparrow  \uparrow  \uparrow  \downarrow } \right\rangle \quad \left| 6 \right\rangle  = \left| { \uparrow  \downarrow  \uparrow  \downarrow } \right\rangle \quad \left| {10} \right\rangle  = \left| { \downarrow  \uparrow  \uparrow  \downarrow } \right\rangle \quad \left| {14} \right\rangle  = \left| { \downarrow  \downarrow  \uparrow  \downarrow } \right\rangle  \\
 \left| 3 \right\rangle  = \left| { \uparrow  \uparrow  \downarrow  \uparrow } \right\rangle \quad \left| 7 \right\rangle  = \left| { \uparrow  \downarrow  \downarrow  \uparrow } \right\rangle \quad \left| {11} \right\rangle  = \left| { \downarrow  \uparrow  \downarrow  \uparrow } \right\rangle \quad \left| {15} \right\rangle  = \left| { \downarrow  \downarrow  \downarrow  \uparrow } \right\rangle  \\
 \left| 4 \right\rangle  = \left| { \uparrow  \uparrow  \downarrow  \downarrow } \right\rangle \quad \left| 8 \right\rangle  = \left| { \uparrow  \downarrow  \downarrow  \downarrow } \right\rangle \quad \left| {12} \right\rangle  = \left| { \downarrow  \uparrow  \downarrow  \downarrow } \right\rangle \quad \left| {16} \right\rangle  = \left| { \downarrow  \downarrow  \downarrow  \downarrow } \right\rangle  \\
 \end{array}
\end{equation}

When nuclear spins are absent ($
\Delta \hat H = 0
$
), Hamiltonian $
\frac{{\hat H_0 }}{J}
$
  has four times four degenerated eigen states (Fig. 2). Two
underlying electron levels have the point of crossing C when  $
\beta  = 1
$
.

Symmetric matrix16*16, corresponding to the matrix of eigen states
of Hamiltonian  $
\Delta \hat h \equiv \frac{{\Delta \hat H}}{J}
$
, in the considered basis has only diagonal elements
$
 - \mu  + \frac{{\alpha _a  + \alpha _b }}{4}
$,
$
 \frac{{\alpha _a  - \alpha _b }}{4}
$,
$
 - \frac{{\alpha _a  - \alpha _b }}{4}
$,
$
  \mu  - \frac{{\alpha _a  + \alpha _b }}{4}
$,
$
 - \mu  + \frac{{\alpha _a  - \alpha _b }}{4}
$,
$
 \frac{{\alpha _a  + \alpha _b }}{4}
$,
$
 - \frac{{\alpha _a  + \alpha _b }}{4}
$,
$
 \mu  - \frac{{\alpha _a  - \alpha _b }}{4}
$,
$
 - \mu  - \frac{{\alpha _a  - \alpha _b }}{4}
$,
$
 - \frac{{\alpha _a  + \alpha _b }}{4}
$,
$
  \frac{{\alpha _a  + \alpha _b }}{4}
$,
$
 \mu  + \frac{{\alpha _a  - \alpha _b }}{4}
$,
$
 - \mu  - \frac{{\alpha _a  + \alpha _b }}{4}
$,
$
 - \frac{{\alpha _a  - \alpha _b }}{4}
$,
$
 \frac{{\alpha _a  - \alpha _b }}{4}
$,
$
 \mu  + \frac{{\alpha _a  + \alpha _b }}{4}
$,
where $
\mu  \equiv \frac{{g_N \mu _N B}}{J},\;\alpha _a  \equiv \frac{{A_a }}{J},\;\alpha _b  \equiv \frac{{A_b }}{J}
$
and non-diagonal elements, from which we point out only up-diagonal elements:

\begin{equation}
\begin{array}{c}
\left( {\Delta \hat h} \right)_{5,2}  = \left( {\Delta \hat h} \right)_{7,4}  = \left( {\Delta \hat h} \right)_{13,10}  = \left( {\Delta \hat h} \right)_{15,12}  = \frac{{\alpha _b }}{2} \\
\left( {\Delta \hat h} \right)_{9,3}  = \left( {\Delta \hat h} \right)_{10,4}  = \left( {\Delta \hat h} \right)_{13,7}  = \left( {\Delta \hat h} \right)_{14,8}  = \frac{{\alpha _a }}{2}
\end{array}
\end{equation}

Since the projection of full electron and nuclear spin in the direction
of the magnetic field $M + m = M_a  + M_b  + m_a  + m_b $  is conserved,
the matrix16*16 breaks down into 5 reduced matrixes of smaller dimension,
corresponding to the values $M + m = 0, \pm 1, \pm 2$
, which diagonalize independently. The next states correspond to
these 5 matrixes:

\begin{equation}
\begin{array}{l}
 m + M = 0:\left| 4 \right\rangle ,\left| 6 \right\rangle ,\left| 7 \right\rangle ,\left| {10} \right\rangle ,\left| {11} \right\rangle ,\left| {13} \right\rangle  \\
 m + M = 1:\left| 2 \right\rangle ,\left| 3 \right\rangle ,\left| 5 \right\rangle ,\left| 9 \right\rangle  \\
 m + M =  - 1:\left| 8 \right\rangle ,\left| {12} \right\rangle ,\left| {14} \right\rangle ,\left| {15} \right\rangle  \\
 m + M = 2:\left| 1 \right\rangle  \\
 m + M =  - 2:\left| {16} \right\rangle . \\
 \end{array}
\end{equation}

With the help of the found matrix, introducing Hamiltonian (\ref{R18}),
the picture of the hyperfine splitting of two donor atoms' electrons' levels
was achieved. It is shown for $
\alpha _a  = 0.3;\quad \alpha _b  = 0.4
$
in figure 3.

The interaction of electronic and nuclear spins not only takes down the
four times degeneracy, but brings near point C the phenomenon of
anticrossing of the levels, which have the same value of $m+M$.

For two crossing electronic levels it takes place for the states with
$m+M=-1$ $|15>$ and $|12>$  (two solid bold lines) and with $m+M=0$ $|10>$
and $|13>$ (two dashed lines) (see Fig. 3).

As the result of this anticrossing the state $|15>$ with $M_a+M_b=-1$
and $m_a+m_b=0$, when J increases (with the help of J-gate), makes a
transition to the $|12>$ state with $m_a+m_b=-1$ and $M_a+M_b=0$, i.e.
the simultaneous inversion of one nuclear spin and one electron spin
takes place (the information about nuclear spin condition is transfered
to the electron spin). Herewith the electron triplet state ($S=1$) makes a
transition to the singlet one ($S=0$).

The energy residual of the underlying anticrossing state $|15>$ and
crossing state $|14>$ in the second order of perturbation theory for the
strong magnetic fields ($
\beta  >  > 1
$
) and for  $
A_a  = A_b  \equiv A
$
, according to \cite{5}, is

\begin{equation} \label{R19}
E_{14}  - E_{15}  = 2\pi \hbar \upsilon _J  = \frac{{\left( {{A \mathord{\left/
 {\vphantom {A 2}} \right.
 \kern-\nulldelimiterspace} 2}} \right)^2 }}{{2\mu _B B - J}} - \frac{{\left( {{A \mathord{\left/
 {\vphantom {A 2}} \right.
 \kern-\nulldelimiterspace} 2}} \right)^2 }}{{2\mu _B B}},
\end{equation}
where $
\upsilon _J
$
 --- transition frequency, which corresponds to projection changing
of two nuclear spins. It increases, when  $
\beta
$
 decreases.

In \cite{5} the levels $|15>$ and $|14>$ are suggested to be used as
"working" ones to conduct the calculation process and to measure the
final state.

However, we also can propose an alternative variant.
It is possible to use the second pair of anticrossing levels
with $m+M=0$. Underlying state of the anticrossing level $|13>$ in the
strong fields corresponds to $M_a+M_b=-1$, $m_a+m_b=1$, and above-lying
state $|14>$ corresponds to $M_a+M_b=-1$, $m_a+m_b=0$, i.e. they differ
by the inverted value of one of the nuclear spins.

When J increases, the state $|13>$, moving through the anticrossing point,
makes a transition to the state $|10>$ with $M_a+M_b=0$ and $m_a+m_b=0$,
i.e. the simultaneous inversion of the electron spin such as the nuclear
one happens (the information about nuclear spin is transferred to the
electron spin, too).

According to \cite{5}, the measurement process of nuclear spin is s
upposed to be implemented in two stages. By changing the value of the
electron exchange interaction J from $
J < 2\mu _B B
$
up to $
J > 2\mu _B B
$
  the adiabatic transition of the nuclear spin condition is made from
triplet electron ground state to the singlet one. Then the electron state
measurement is made.

If at first the electrons of two donors are at the triplet state,
simultaneously being placed on its own atom, then, when J increases,
it is more energy-advantageous for the both electrons to exist in the
singlet state being placed together on the one of the donors, if the bonding
energy of two electrons on the single donor exceeds the bonding energy for
each electron on its own donor ($
D^ -
$
 -state). As a result the charge carrying takes place from one donor
to another. As it is pointed out in \cite{5}, this fact can be registered
by highly sensitive single-electron capacity methods.

At the present time different methods of the spin states measurement
are discussed. In \cite{11} it is suggested to use a scheme, agreeably
called "turnstile", which omits current of electrons only with the
definite spin state. This electric method allows to measure electron
spin states. Another approach was discussed in \cite{12}. It is based on
using magnetic resonant force microscopy.

\section {Conclusion}

In this work an influence of the electric field, inducing by the A-gate
(for the gate as a round disc and as a strap) on the hyperfine interaction
constant was analyzed in details. The weaker dependence of the hyperfine
interaction constant from the depth of underlying donors under the gate,
and also decreasing of its magnitude when the distance from the gate to
the substrate increases are noted. Conducted calculations with two gate
types allows us to make a conclusion about significant dependence of the
hyperfine interaction constant from the form of a gate.

Parameters a and c (5-10 nm) of the structure at which errors rate can
be minimized are determined by (17). Obtained expressions allow us to
determine admissible values of technological inaccuracies $\delta x$
 and $\delta z$
 (2-3 nm) when the constant A errors and working potential
values V (0.1-1 volts) of a gate are defined.
In work the full energy spectrum of the electron-nuclear spin system of
two interacting donor atoms for $A_a  \ne A_b $
 was calculated.
The existence of two pairs of anticrossing levels at the ground electron
state, corresponding to the full electron and nuclear spin
projection  $m_a+m_b+M_a+M_b=-1$ and  $m_a+m_b+M_a+M_b=0$ was demonstrated.
Underlying components of these pairs correspond to the transition from
the triplet electron state (S=1) to the singlet one (S=0), when the field
decreases. If the full electron spin projection changes
from $M_a+M_b=-1$ to $M_a+M_b=0$, the nuclear spin projection also
correspondingly changes. Thus the information about nuclear spin state
transfers to the electron system.
Both pairs of anticrossing levels may be used to conduct calculation and
measurement process. Herewith we must use the frequency, determined by
splitting between underlying anticrossing levels ($|13>$ or $|15>$) and a
bove-lying level $|14>$, to change nuclear spin state.

In conclusion authors would like to thank L.A. Openov for the valuable
comments.

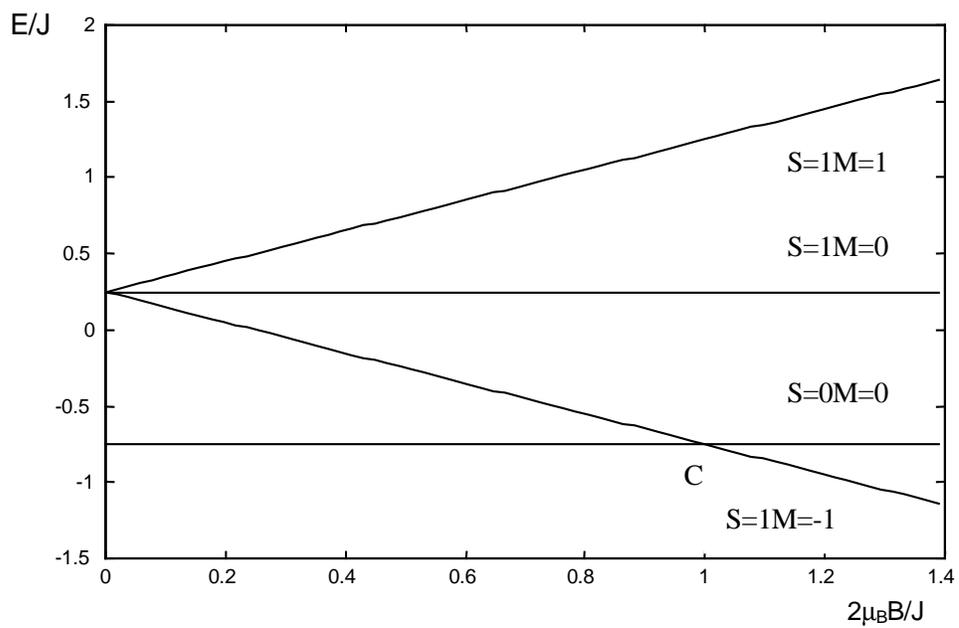
\end{document}